\documentclass{PoS}
\newcommand{\Ef}{\mathcal{E}}
\newcommand{\Ea}{\mbox{E}_{300}}
\newcommand{\Eb}{\mbox{E}^L_{300}}
\newcommand{\Ec}{\mbox{E}_{220}}
\usepackage{amsmath}
\title{Valence calculation of the electric polarizability on nHYP-Clover ensembles.}

\ShortTitle{Valence calculation of the electric polarizability}

\author{\speaker{Michael Lujan},~~Andrei Alexandru, Walter Freeman,~and~Frank Lee\\
        The George Washington University, Washington DC, USA\\
        E-mail: \email{mlujan@gwmail.gwu.edu},~~ \email{aalexan@gwu.edu},~~ \email{wfreeman@gwu.edu},~~ \email{fxlee@gwu.edu}}

\abstract{We present preliminary calculations for the electric polarizability of the neutral pion and neutron on three dynamically generated nHYP-Clover ensembles. We use two different pion masses ($m_{\pi} \simeq 300$ and 220 MeV) to gauge the chiral behavior.
The effects of partial quenching are analyzed by computing a string of partial quenched valence masses for each ensemble. We also analyzed the volume dependence using elongated lattices, where the elongation is in the direction of the electric field.    
} 

\FullConference{31st International Symposium on Lattice Field Theory - LATTICE 2013\\
		July 29 - August 3, 2013\\
		Mainz, Germany}

\begin{document}

\section{Introduction}
At lowest order the effects of an electric field, $\Ef$, and magnetic field, $B$, on hadrons can be parameterized by the effective Hamiltonian:
\begin{equation}
\mathcal{H}_{em} = -\vec{p}\cdot\vec{\Ef} -\vec{\mu}\cdot\vec{B} -\frac{1}{2}\left(\alpha \Ef^2 + \beta B^2\right)+...,
\end{equation}
where $p$ and $\mu$ are the static electric and magnetic dipole moments, respectively, and $\alpha$ and $\beta$ are the electric and magnetic polarizabilities. Due to time reversal symmetry of the strong interaction, the static dipole moment, $p$, vanishes.  Furthermore, by restricting ourselves to the case of a constant electric field, the leading contribution to the electromagnetic interaction comes from the electric polarizability term at $\mathcal{O}(\Ef^2)$. The effect of the external field to the ground state is observed as an overall energy shift {\it{i.e.}}

\begin{equation}
E = E_0 -\frac{1}{2} \alpha \Ef^2 + ... .
\end{equation}
Thus, calculating the energy shift leads us directly to the polarizability. 

On the lattice the first step to compute $\alpha$ is to introduce the electric field. Our work employs the commonly used background field method \cite{Martinelli:1982cb}.  This procedure introduces the electromagnetic vector potential, $A_{\mu}$, to the Euclidean QCD Lagrangian via minimal coupling; the covariant derivative becomes
\begin{equation}
D_{\mu} = \partial_{\mu} -igG_{\mu} -iqA_{\mu},
\end{equation}
where $G_{\mu}$ are the gluon field degrees of freedom. The practical implementation of  this is achieved by a multiplicative phase factor to the gauge links {\it{i.e.}}
\begin{equation}
U_{\mu} \rightarrow e^{-iqaA_{\mu}} U_{\mu}.
\end{equation}

To obtain a constant electric field, say in the $x$-direction, we may choose $A_{x} = -i\Ef t$ which provides us with a real phase factor.  
A more convenient choice, and the one implemented in this study,  is the use of an imaginary value for the electric field so that links remain unitary. For a more complete discussion on the use of real or imaginary fields see \cite{Alexandru:2008sj}.

After placing the field onto the lattice we calculate the zero-field ($G_0$), plus-field ($G_{+\Ef}$),  and minus-field ($G_{-\Ef}$) two-point correlation functions for the interpolating operators of interest. In this work we will focus on the neutral pion and neturon. For the neutral pion we did not include the disconnected contributions which are required since the quarks' charge breaks the isospin symmetry.  The combination of the plus and minus field correlators allows us to remove any $\mathcal{O}(\Ef)$ effects, which are statistical artifacts. For neutral particles in a constant electric field the correlation functions still retain their single exponential decay in the limit $t \rightarrow \infty$. In particular we have
\begin{equation}
\lim_{t \to \infty} G_{\Ef} = A(\Ef)\exp[-E(\Ef)~t], 
\label{eqn::corr}
\end{equation}
where $E(\Ef)$ has the perturbative expansion in the electric field given by
\begin{equation}
E(\Ef) = m + \frac{1}{2}\alpha \Ef^2 + ... ~.
\end{equation}
By studying the variation of the correlation functions with and without an electric field one can isolate the energy shift to obtain $\alpha$.

For the case of spin 1/2 particles the energy $E(\Ef)$ is altered because of the fact that magnetic moment of the particle also enters at $\mathcal{O}(\Ef^2)$ \cite{Lvov:1993fp}. The form of the correlation function is still given by Eqn.~\ref{eqn::corr} but with 
\begin{equation}
E(\Ef) = m + \frac{1}{2}\Ef^2\left(\bar{\alpha}- \frac{\mu^2}{m}\right) + ...~,
\label{eqn::Eneutron}
\end{equation}
where $\bar{\alpha}$ is the Compton polarizability. Thus $\bar{\alpha}$ is related to the static polarizability, $\alpha$, via
\begin{equation}
 \bar{\alpha} = \alpha + \mu^2/m.
\end{equation}
In this work we will assume for now that the magnetic moment contribution is given approximately by the continuum value, $\mu = -1.9 \mu_N$ where $\mu_N$ is the nuclear magneton.

\section{Calculation Details}
\subsection{Choice of the Electric Field and Boundary Conditions}
\begin{center}
\begin{figure*}[t]
\includegraphics[width= 3.2in]{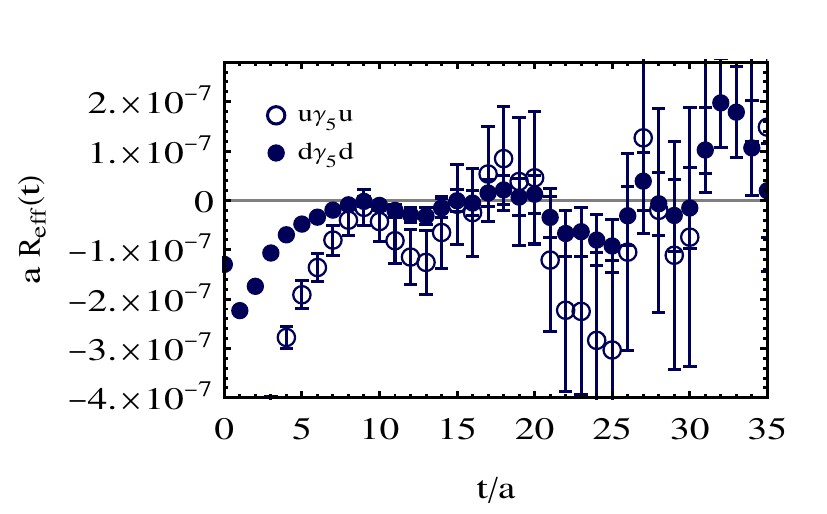}
\includegraphics[width= 3.2in]{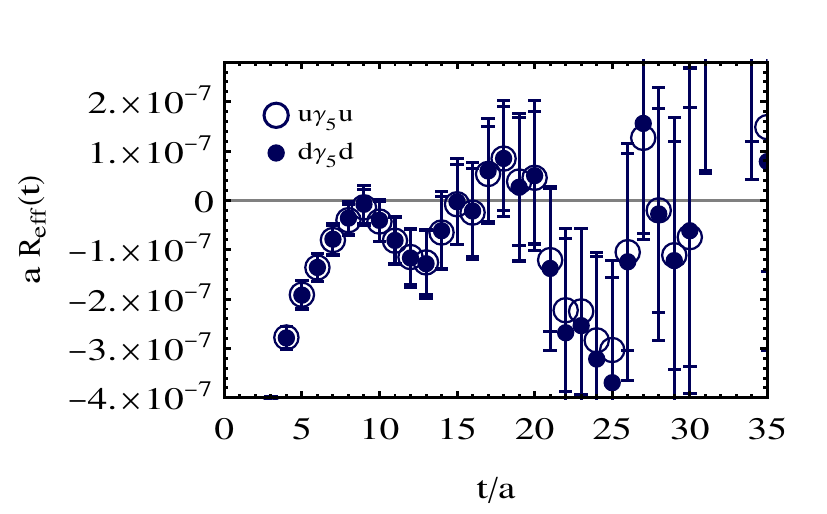}
\caption{Left panel: The ratio of effective mass plot for the interpolators $\bar{u}\gamma_5 u$ and $\bar{d}\gamma_5 d$ for the $\Ea$ ensemble~(see Table~1). The right panel is the same plot but with the values of $\bar{d}\gamma_5 d$ scaled by a factor of 4.} 
\label{plot:fieldstrength}
\end{figure*}
\end{center}
The extraction of the polarizability relies on the assumption that we are using perturbatively small values of the electric field. Our calculations have been using a field value of $\eta~\equiv a^2 q_d \Ef =~10^{-4}$, where $q_d$ is the charge of the down quark. We can determine if this field strength is in the perturbative regime by looking at the effect of the electric field on the interpolators $\bar{u}\gamma_5 u$ and $\bar{d}\gamma_5 d$ separately. We expect that the energy shift of $\bar{d}\gamma_5 d$ be four times smaller than $\bar{u}\gamma_5 u$ because its charge is twice as small. To study this we compute the effective mass of the ratio between the zero field and electric field correlators given by
\begin{equation}
R(t) = \frac{G_{\Ef}(t)}{G_{0}(t)}  \overset{t\rightarrow \infty}{\rightarrow} e^{-\Delta E t}.
\end{equation}
Our results are presented in Fig.~\ref{plot:fieldstrength}. On the right panel, of the same figure, we plot the same thing but with the the values of $\bar{d}\gamma_5 d$ scaled by a factor of 4.  If we are in the region where $\mathcal{O}(\Ef^2)$ contributions dominate then the two effective masses should agree very precisely with the scaled $\bar{d}\gamma_5 d$ effective mass. This behavior is indeed what we see and gives us confidence that we are using a suitable value for the field.

\begin{figure}[t]
\centering
\includegraphics[width= 3.4in]{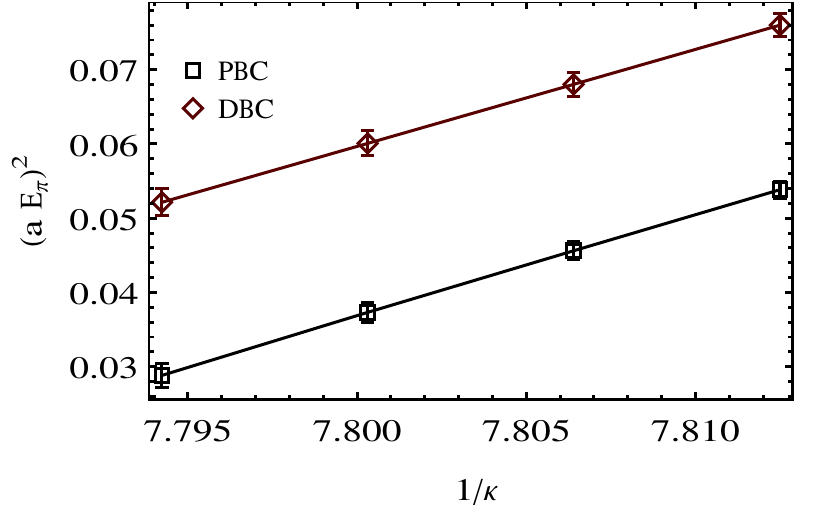}
\caption{Plot of $(a E_{\pi})^2$ as a function of $1/\kappa$ for Dirichlet and periodic boundary conditions}
\label{plot:pionbc}
\end{figure}

The use of the relatively small value of the field is made possible because of the fact that we use Dirichlet boundary conditions (DBC) in both the time and direction of the field. If we were to use periodic boundary conditions we would have to quantize the value of field in units of $\Ef_0 = 2 \pi/(q T L)$ where $T$ and $L$ are the physical extent of the lattice in the time and spatial directions respectively. These values are much larger and we do not know whether or not this field value is small enough to neglect higher order effects.  Using DBC does not constrain us to use these quantized values, they do however introduce boundary effects. One such effect is that it induces a non-zero momentum of order $\pi/L$ which vanishes only in the limit $L \rightarrow \infty$. This additional momentum alters the energy state of the hadron. One can see this by looking at the lowest energy in the pion channel with DBC and PBC.  This is shown in Fig.~\ref{plot:pionbc}.  
When the hadron is moving the energy shift, $\delta E$, induced by the electric field is not equal to the change in the hadron mass since $E = \sqrt{m^2 + p^2}$. We compute the mass shift, $\delta m$, as
\begin{equation}
\delta m = \delta E \frac{E}{m},
\end{equation}
where $m$ is the zero-momentum mass of the particle which we calculate using PBC.

\subsection{Ensemble Details}
\begin{figure*}[t]
\includegraphics[width= 2.7in]{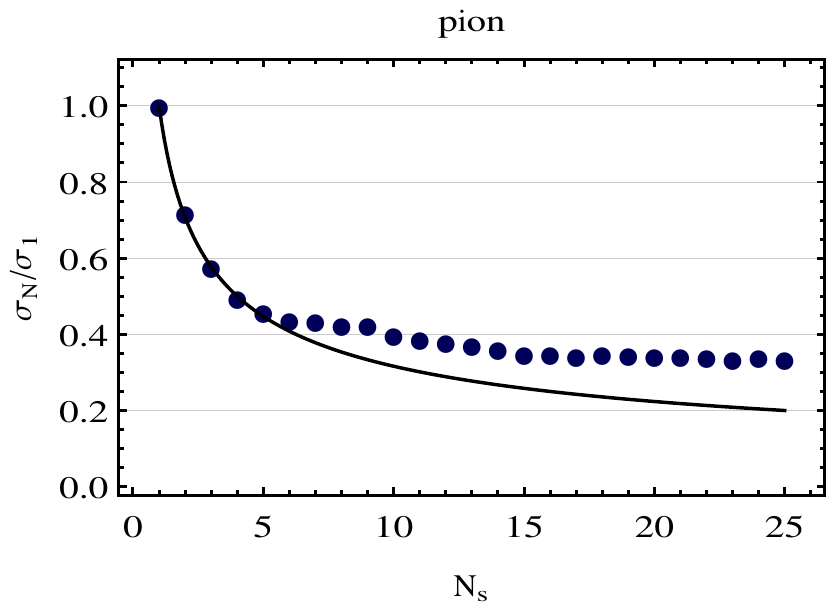}
\hspace{0.2 in}
\includegraphics[width= 2.7in]{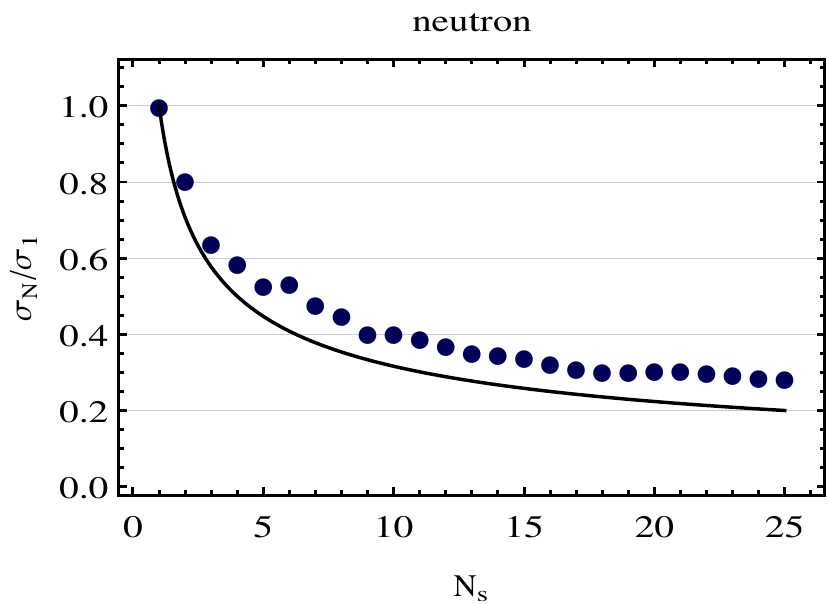}
\caption{Scaling of the pion and neutron. The blue points correspond to the measured values of the uncertainty in the energy shift, $\delta E$. The solid black line corresponds to the expected ideal situation if the data points were completely uncorrelated. The curves were generated by taking the value for $N_s=1$ and scaling it by $1/\sqrt{N_s}$.}
\label{plot:sources}
\end{figure*}

We computed the polarizability on three dynamically generated nHYP-Clover ensembles. The details of the ensembles are listed in Table~\ref{tab:ensembles}. 
\begin{table}[b]
\begin{center}
\begin{tabular}{|c|c|c|c|c|c|}
\hline
Ensemble&Lattice Size&  $a$ (fm) &  $\kappa$& $m_{\pi}$ (MeV)& $N_{\mbox{configs}}$\\ 
\hline
\hline
$\Ea$&$24\times 24^2 \times 48$& 0.1255(7)& 0.1282   & $\simeq 300$& 300\\
$\Eb$&$48\times 24^2 \times 48$& 0.1255(7)& 0.1282   & $\simeq 300$& 300\\
$\Ec$&$24\times 24^2 \times 64$& 0.1255(7)& 0.12838 & $\simeq 220$& 500\\
\hline
\end{tabular}
\end{center}
\caption{Details of the lattice ensembles used in this work. The lattice spacing was determined in \cite{Pelissier:2012pi}. }
\label{tab:ensembles}
\end{table}
For each ensemble we computed roughly 20 sources per configuration. The different sources are of course correlated; at some point we expect the uncertainty in the extracted parameters to plateau as we increase the number of sources. The benefits of using multiple sources are shown in Fig.~\ref{plot:sources}. There we plot the value of the uncertainty in the energy shift as a function of the number of sources for the $\Ea$ ensemble. We see that for the pion not much is gained by using more than $\sim$ 15 sources, however the neutron still benefits from using all the sources. A plateau behavior, similar to the pion case, is expected for the neutron if we were to increase the number of sources further. 

Along with the dynamical pion masses we computed correlators for a string of partially quenched pion masses.  The partially quenched $m_{\pi}$ values for the $\Ea$ and $\Eb$ ensembles are: $m_{\pi} = 270$, 340 and 370 MeV. ~For the $\Ec$ ensemble we used similar values: $m_{\pi} =$ 270, 300, 340 and 370 MeV. To increase our efficiency of quark propagator calculations we used an optimally implemented multi-GPU Dslash operator \cite{Alexandru:2011sc}, along with an efficient BiCGstab multi-mass inverter~\cite{Alexandru:2011ee}.

\section{Results}
Extracting the energy shift, $\delta E$, induced by the electric field requires a simultaneous-correlated fit to the zero, plus, and minus field correlation functions. The correlation functions are highly correlated because they come from the same gauge configurations and differ only by the perturbatively small electric field. We emphasize that the desired energy shift is very small, several orders of magnitude smaller than the statistical uncertainties of the energy itself. Thus, our only hope of extracting such a tiny shift lies in the ability to properly account for the correlations among the three values of the electric field. To do this we construct the following difference vector as

\begin{equation}
\mathbf{v}(t,\eta) \equiv (A +\delta A~\eta^2) e^{-(E + \Delta E\eta^2)t}.
\end{equation}
We minimize $\chi^2$ in the usual manner with the correlation matrix given by
\[ \mathbf{C} = \left( \begin{array}{ccc}
C_{0 0} ~& C_{0 +}~ & C_{0 -} \\
C_{+ 0} ~& C_{+ +}~ & C_{+ -} \\
C_{- 0} ~& C_{- +} ~& C_{- -} \end{array} \right),\] 
where $0,+,-$ represent  $G_0, G_{+\Ef}$, and $G_{-\Ef}$ respectively. 

\begin{figure*}[t]
\includegraphics[width= 3.2in]{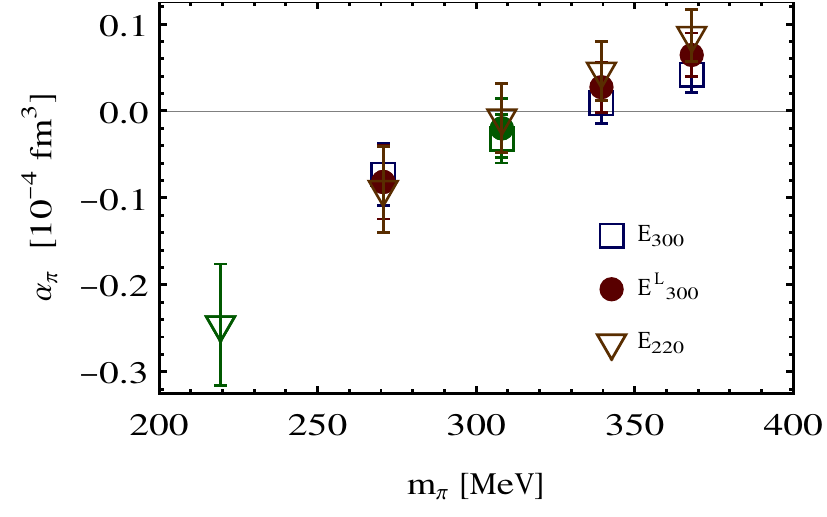}
\includegraphics[width= 3.3in]{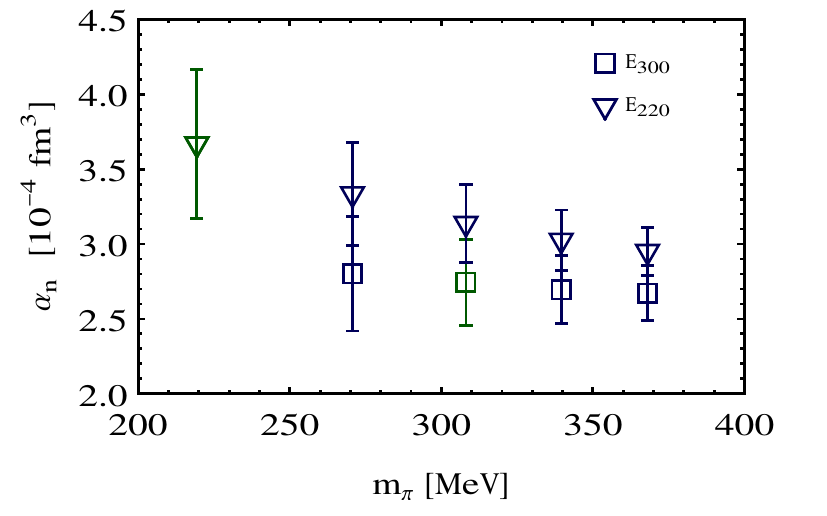}
\caption{Preliminary results for the pion and neutron polarizability. The Green symbols highlight the {\em{dynamical}} points for each lattice, i.e. the points where the valence mass is equal to the sea mass.}
\label{plot:results}
\end{figure*}

Our preliminary results for the pion and neutron are shown in Fig.~\ref{plot:results}. The pion data seem to show very little or no dependence on the volume or the sea quark mass.  The results also show the same negative trend that has been observed in other studies \cite{Alexandru:2008sj,Detmold:2009dx}. It was suggested that perhaps it was due to finite volume effects,  however, the study done in \cite{Alexandru:2010dx} along with our study show that this is not the case. It is possible that this effect is due to the fact that we have left out the contribution of the charge of the sea quarks. We are investigating this possibility \cite{Freeman:2012cy}. 

The neutron polarizability indicates a promising rise for the $\Ec$ ensemble. However, preliminary comparisons suggest that predictions from $\chi$PT are still well above the values we are currently seeing for the neutron polarizabilities. Our data $\Eb$ ensemble have very large error bars as of now and thus we do not yet present those results. We are currently generating more sources to reduce the statistical error. It may very well be that finite volume effects are responsible for the discrepancy from the $\chi$PT predictions.
\section{Conclusion and Outlook}
We performed a volume and chiral study for the pion and neutron polarizabilities using nHYP Clover fermions at two dynamical pion masses: 300 MeV and 220 MeV. These are to date the lowest pion masses used in polarizability studies. We also performed a volume analysis using elongated lattices where the elongation is in the direction of the electric field. The value of the pion polarizability becomes more negative as we decrease the pion mass which is consistent with what has been observed in other works. The neutron polarizability for the $\Ec$ ensemble shows a rise towards the physical point as a function of the pion mass. We have not yet presented our results for the volume study for the neutron. We are currently computing more sources to reduce the statistical uncertainties in our data.  A parallel project is also being done to include the effects of the sea quarks. This work is explained in  \cite{Freeman:2012cy}  and is being done on the same $\Ea$ ensemble used here. 

\section{Acknowledgements} 
We would like to thank Craig Pelissier for generating all the ensembles used in this study. This work was done on the following GPU clusters: GWU IMPACT clusters, GWU CCAS Colonial One cluster, JLab clusters,   Fermilab clusters, and UK clusters.  This work is supported in part by the NSF CAREER grant PHY-1151648 and the U.S. Department of Energy grant DE-FG02-95ER-40907.

\bibliographystyle{jhep-3}
\bibliography{my-references}

\end{document}